\documentstyle[sprocl]{article}

\def\Journal#1#2#3#4{{#1} {\bf #2}, #3 (#4)}

\def\NPB{{\em Nucl. Phys.} B}
\def\PLB{{\em Phys. Lett.}  B}
\def\PRL{\em Phys. Rev. Lett.}
\def\PRD{{\em Phys. Rev.} D}

\newcommand{\be}{\begin{equation}}
\newcommand{\ee}{\end{equation}}
\newcommand{\bea}{\begin{eqnarray}}
\newcommand{\eea}{\end{eqnarray}}

\newcommand{\p}[1]{(\ref{#1})}

\begin{document} 
\begin{flushright}
Preprint DFPD 97/TH/41\\ 
hep-th/9709190 \\
September, 1997 
\end{flushright}

\title{SYMMETRY PROPERTIES OF SELF--DUAL FIELDS\footnote{
Talk given at the Fifth International 
Wigner Symposium (Vienna, August 25-29, 1997)
}}

\author{ D. P. SOROKIN\footnote{Permanent address: NSC,
Kharkov Institute of Physics and Technology,
Kharkov, 310108, Ukraine; sorokin@pd.infn.it dsorokin@kipt.kharkov.ua}}

\address{Universit\`a Degli Studi Di Padova,
Dipartimento Di Fisica ``Galileo Galilei''\\
ed INFN, Sezione Di Padova,
Via F. Marzolo, 8, 35131 Padova, Italia}

\maketitle\abstracts{
We briefly review the structure and properties of self-dual field 
actions.}

The Lagrangian description of gauge fields which possess duality 
properties have been under study during a long period of time, a 
classical example being 
the electric--magnetic duality symmetry of the free 
Maxwell equations. 

Present interest in this problem is caused by an essential progress in 
understanding an important role of dualities in quantum string theory and 
its effective field--theoretical limits.
Self--dual and duality--symmetric fields are part of the spectrum in 
these theories and the knowledge of their effective actions would enable 
one to get more detailed information about the dynamical properties, 
symmetries and geometrical structure of the theory. And here a problem of 
how to construct a duality symmetric action arises. On a way of solving 
this problem one observes that self--dual (or duality--symmetric) fields 
have unusual group--theoretical properties and non--trivial topological 
structure inherent to this important subclass of gauge fields.

I shall describe this properties with the example of a self--dual
antisymmetric 2--rank field $A_{mn}(x)$ in 6-dimensional 
space--time (m,n = 0,...,5), though an approach \cite{pst} to be used is 
applicable to all known models with duality--symmetric fields in any 
space--time dimension of Lorentz signature.

A field $A_{mn}(x)$ is self--dual if  
\be\label{1}
\partial_{[l}A_{mn]}(x)\equiv
F_{lmn}={1\over 6}\epsilon_{lmnpqr}F^{pqr}\equiv F^*_{lmn}.
\ee
This additional condition on the field strength of $A_{mn}$ reduces twice 
the number of physical degrees of freedom of the gauge field which 
otherwise would have 6 transversal degrees of freedom. 
Upon imposing \p{1} $A_{mn}$ has only three 
independent degrees of freedom. It is desirable 
to describe the dynamics of $A_{mn}$ by 
use of the action principle, and consider \p{1} as the equation of motion 
of $A_{mn}$ derived from an action. 
Since \p{1} 
is a first order differential equation one might admit that the action 
should be of the first order in derivatives. This is not typical of the 
bosonic fields whose free actions are usually quadratic in field 
derivatives (in a second-order formalism). If anyway we try to 
construct a first--order action, we shall realize that an infinite number 
of auxiliary fields are required \cite{mwy} for the action to consistently 
describe a single self--dual field. 

Another possibility is looking for a quadratic action constructed solely 
from components of $F_{lmn}$. Then, since equations of motion one 
derives from a quadratic action are of the second order in derivatives, 
one may expect that there should be additional local symmetry which 
would allow one to reduce these equations to the self--duality 
condition. Such actions have indeed been found for various models 
\cite{z,fj} but they turn out to be nonmanifestly space--time invariant.
In our case the self--dual action can be written as
\be\label{2}
S=\int d^6x [-{1\over 6}F_{lmn}F^{lmn}
-{1\over{2}}(F-F^*)_{0ab}
(F-F^*)^{~ab}_0], \qquad (a,b = 1,...,5)
\ee
It contains the ordinary kinetic term for $A_{mn}$, and the second term 
which breaks manifest Lorentz invariance down to its spatial 
subgroup $SO(5)$, since only time 
components of $(F-F^*)$ enter the action. 
However, it turns out 
that Eq. \p{2} is non--manifestly invariant under modified 
space--time transformations \cite{fj} which (in the gauge
$A_{0i}=0$) look like
\be\label{3}
\delta A_{ab}=x^0v^c\partial_cA_{ab}+x^cv^c\partial_0A_{ab}-
x^cv^c(F-F^*)_{0ab}.
\ee
The first two terms in \p{3} are standard Lorentz boosts with a velocity
$v_a$, and the last term is a nonconventional one, it vanishes when
\p{1} is satisfied, so that the transformations \p{3} reduce to the 
Lorentz boosts on the mass shell.\\
From \p{2} we get $A_{mn}$ field equations, which have the 
form of Bianchi identities
$
\epsilon^{abcde}\partial_a(F-F^*)_{bc0}=0.
$
Their general (topologically trivial) solution is
\be\label{5}
(F-F^*)_{ab0}=\partial_{[a}\varphi_{b]}(x).
\ee
If the right hand side of \p{5} was zero then $F_{ab0}=F^*_{ab0}$
and, hence, as one can easily check, 
the full covariant self--duality condition would be satisfied. 
And this is what we 
would like to get. One could put the r.h.s. of \p{5} to zero if there 
would be an additional local symmetry of \p{2} for which 
$\partial_{[a}\varphi_{b]}=0$ is a gauge fixing condition. And there is 
indeed such a symmetry \cite{} under
\be\label{6}
\delta A_{0a}=\hat\varphi_a(x),\qquad \delta A_{ab}=0, \qquad
\delta(F-F^*)_{ab0}=\partial_{[a}\hat\varphi_{b]}.
\ee
The existence of this symmetry is the reason why the quadratic action 
describes the dynamics of the self--dual field $A_{mn}$ with twice less 
physical degrees of freedom than that of a non--self--dual one.

Thus we have constructed a non--manifestly Lorentz invariant action
for the self--dual fields.
But it is always desirable to have a covariant formulation, especially 
when one intends to consider more complicated models coupled to 
(super)gravity.
This requires the use of auxiliary fields. Depending on the
approach chosen their number vary from one \cite{pst} to \cite{si} 
infinity \cite{mwy}.

I will briefly describe a formulation with a 
single scalar auxiliary field $a(x)$ which has been used
to construct the effective action for the M--theory 5--brane 
\cite{m}. The covariant action for the $D=6$ self--dual field looks as 
follows  
\begin{equation}\label{fina}
S=\int d^6x [-{1\over 6}F_{lmn}F^{lmn}
+{1\over{2(\partial_qa\partial^qa)}}\partial^ma(x)(F-F^*)_{mnl}
(F-F^*)^{nlr}\partial_ra(x)].
\end{equation}
 In addition to ordinary gauge symmetry of $A_{mn}(x)$ the 
covariant action 
(\ref{fina}) is invariant under the following local  
transformations:
\begin{equation}\label{varm}
\delta A_{mn} =\partial_{[m}a\hat\varphi_{n]}(x), \qquad
\delta a=0;
\end{equation}
\begin{equation}\label{a}
\delta a=\varphi(x), \qquad
\delta A_{mn}=
{{\varphi(x)}\over{(\partial a)^2}}{\cal F}_{mnp}\partial a^p. 
\end{equation}
The transformations \p{varm} are a covariant counterpart of \p{6} and
play the same role as the latter in deriving the self--duality 
condition \p{1}.

Local symmetry \p{a} ensures the auxiliary nature of 
the field $a(x)$ required for keeping space--time 
covariance of the action manifest \cite{pst}. 
An admissible gauge fixing condition 
$
\partial_m a(x)=\delta^0_m
$
for this symmetry
reduces \p{fina} to \p{2}, the modified
space--time transformations \p{3}, which
preserve this gauge, arising as a combination of the 
Lorentz boost and the transformation \p{a} with $\varphi=-v^cx^c$.

We have thus seen that the self--dual gauge fields possess more local 
symmetries than the ordinary gauge fields, and a nonpolinomial form of
the covariant action \p{fina}, which is singular at $(\partial a)^2=0$,
points to a nontrivial topological structure associated with these 
fields. Better understanding the nature of these symmetries and of the 
topological structure may be useful for studying quantum 
self--dual field theory.

\end{document}